# Elder-Sim: A Psychometrically Validated Platform for Personality-Stable Elderly Digital Twins

Jiaqing Wang[1,2], Zhongfang Yang[3], Xingyuan Zhu[4], Zong'an Huang[2], Hao Wang[5], Li Tian[3,6], Ying Cao[7], Xiaomin Qu[8], Xiang Qi[9], Bei Wu[10], Zheng Zhu[1,2]

[1]  Yulin AI-Enhanced HealthCare Lab, Shanghai, China

[2]  School of Nursing, Fudan University, Shanghai, China

[3]  School of Nursing, Medical College of Soochow University, Suzhou, Jiangsu Province, China

[4]  School of Nursing, Dali University, Dali, Yunnan Province, China

[5]  Department of Advanced Interdisciplinary Studies, The University of Tokyo, Japan

[6]  The First Affiliated Hospital of Soochow University, Suzhou, Jiangsu Province, China

[7]  Department of Nursing, The First Affiliated Hospital, Jiangxi Medical College, Nanchang University, Nanchang, Jiangxi Province, China

[8]  School of Social Development, East China University of Political Science and Law, Shanghai, China

[9]  Rory Meyers College of Nursing, New York University, NYC, US

[10] New York University Shanghai, Shanghai, China

## Abstract

**Background:** Large language models (LLMs) enable patient-facing conversational agents, creating a plausible pathway toward patient digital twins that can capture older adults' lived experiences, beliefs, and behavioral responses across time. A central barrier to clinical-grade digital twins is personality drift—inconsistent trait expression across repeated, longitudinal interactions—which can undermine the reliability of generated trajectories and intervention-response simulation in geriatric care contexts.

**Objective:** To develop ELDER-SIM as a multi-role elderly-care conversational platform oriented toward building personality-stable elderly digital twin agents, and to propose a psychometric validation framework for quantifying and improving personality consistency in LLM-based agents across repeated interactions.

**Methods:** ELDER-SIM was implemented using n8n workflow orchestration with local LLM inference (Ollama/vLLM), integrating (1) Big Five (OCEAN) trait specifications, (2) a three-layer Cognitive Conceptualization Diagram (CCD) grounded in Beck's cognitive behavioral therapy framework, and (3) a MySQL-based long-term memory module for persistent states. We conducted a systematic ablation study across four conditions—Baseline (prompt-only), +Memory, +CCD, and +LoRA (domain-specific fine-tuning using 19,717 instruction pairs derived from CHARLS)—and evaluated personality consistency using Cronbach's α (internal consistency), intraclass correlation coefficient (ICC) (test–retest reliability), and role discrimination accuracy .

**Results:** Personality measurement reliability was acceptable to excellent across conditions (Cronbach's α: 0.70–0.94), with consistently high test–retest stability (ICC: 0.85–0.96). Role discrimination improved stepwise from 83.3% (Baseline) to 88.9% (+Memory), 94.4% (+CCD), and 97.2% (+LoRA). CCD produced the largest gain in internal consistency (mean α 0.702→0.892), while LoRA achieved the highest overall internal consistency (α 0.940) and ICC (0.958).

**Conclusions:** ELDER-SIM provides a psychometrically validated approach for constructing personality-consistent elderly digital twin agents. By demonstrating that structured cognitive modeling and domain adaptation materially reduce personality drift, the framework supports more reliable longitudinal digital twins for elderly mental health and psychosocial care, enabling reproducible in silico evaluation of interaction trajectories and intervention strategies before clinical deployment.

# 1. Introduction

## 1.1 Background and Rationale

Population ageing is accelerating globally: the number of people aged ≥60 years is expected to increase from about 1.0 billion in 2020 to 1.4 billion by 2030 and to 2.1 billion by 2050, creating sustained pressure on clinical care delivery and health-system capacity [1]. As older adults live longer with multimorbidity and functional vulnerability, the clinical challenge is increasingly longitudinal: care must anticipate evolving needs, support self-management, and preserve quality of life over time rather than address isolated episodes of illness [2,3]. Within this trajectory, mental health and psychosocial well-being are not secondary concerns but core determinants of outcomes. Depressive and anxiety symptoms, loneliness, and social isolation are common in later life and are associated with poorer quality of life, greater symptom burden, reduced adherence, and higher healthcare utilization [4]. In addition, ageism has been recognized as a modifiable social determinant; perceived age discrimination is consistently linked to psychological distress and adverse health outcomes and can erode trust and engagement in care [5–7]. Clinicians therefore require tools that can account for both biomedical complexity and the stable person-level factors that shape behavior and response to interventions across repeated encounters.

Digital health technologies are expanding, yet a persistent gap is the inability to model, in a clinically meaningful way, how an individual older adult's behavior and psychosocial responses evolve while remaining anchored to relatively stable traits such as personality, beliefs, and coping patterns. Patient digital twins—computational representations designed to mirror an individual's states and trajectories—offer a potential route to personalized planning, scenario testing, and intervention optimization for older adults [8,9]. However, conventional approaches used to approximate patient behavior (eg, scripted vignettes, standardized encounters, rule-based simulators) are limited in scalability and often fail to reproduce the nuanced, context-sensitive, time-dependent interactions characteristic of geriatric psychosocial care [10]. Recent progress in large language models (LLMs) has renewed interest in conversational digital twins because LLM agents can generate rich narratives, sustain dialogue, and incorporate contextual information in ways that better approximate real clinical conversations, with early applications emerging across communication support, decision support, and health education [11–18]. For a digital twin to be clinically credible, however, it must be longitudinally reliable: it should not only produce plausible responses in a single session but also maintain stable identity-consistent patterns across repeated interactions so that outputs can support inference and decision-making.

A major unresolved barrier is personality drift, whereby LLM-based agents shift behavioral tendencies across multi-turn or repeated conversations even when a persona is specified, undermining reproducibility and threatening the validity of any longitudinal trajectory or intervention-response modelling [19–21]. Existing evaluations of LLM health agents have largely focused on task performance and safety, while rarely applying psychometric approaches to test whether intended personality attributes are expressed consistently and can be measured with acceptable reliability [21–23]. This is particularly salient in geriatric psychosocial care, where behavior is mediated by structured cognition—links between life history, core beliefs, intermediate beliefs, coping strategies, and situation-triggered automatic thoughts—relationships formalized within cognitive behavioral therapy frameworks but seldom operationalized in LLM agent construction [22,24,25]. Moreover, general-purpose LLMs may not reflect domain-specific behavioral distributions required for elderly care, motivating domain adaptation approaches such as low-rank adaptation (LoRA), yet their contribution to improving longitudinal identity stability remains insufficiently characterized [17,26]. Collectively, these gaps indicate that progress toward clinically usable elderly digital twins requires (1) architectures that constrain identity and cognition over time and (2) validation frameworks that quantify stability with psychometric rigor.

Against this backdrop, we developed ELDER-SIM as a multi-role elderly-care agent platform explicitly oriented toward constructing personality-stable elderly digital twins rather than single-session chatbots. The study aimed to (i) design an identity-constrained architecture that integrates Big Five (OCEAN) trait specifications with a cognitive-behavioral conceptualization framework to stabilize trait-consistent reasoning across longitudinal interactions; (ii) incorporate persistent memory and domain adaptation to support clinically plausible continuity in older-adult narratives and behavioral responses; and (iii) establish a psychometric evaluation framework to quantify personality stability—using internal consistency, test–retest reliability, and role discrimination—across repeated scenario-based encounters relevant to geriatric psychosocial care. By providing both an implementable platform and a validation protocol, we sought to define reproducible standards for assessing whether LLM-based elderly digital twins can maintain stable person-level characteristics required for longitudinal modelling and in silico testing of intervention strategies.

# 2. Methods

## 2.1 Study Design Overview

The study included four prespecified phases. (1) We designed and implemented a modular platform architecture integrating workflow orchestration, local LLM inference, personality modeling, and persistent memory to support diverse elderly care scenarios. (2) We operationalized agent identity using structured personality specifications and cognitive modeling to promote stable, trait-consistent responses over time. (3) We evaluated personality consistency using a standardized scenario battery with repeated runs and prespecified psychometric indices, including internal consistency (Cronbach α), test-retest reliability (intraclass correlation coefficient [ICC]), and role discrimination accuracy. (4) We conducted controlled ablation experiments to quantify the incremental contribution of key system components (memory, cognitive conceptualization diagram [CCD], and LoRA) to personality stability under otherwise identical testing conditions.

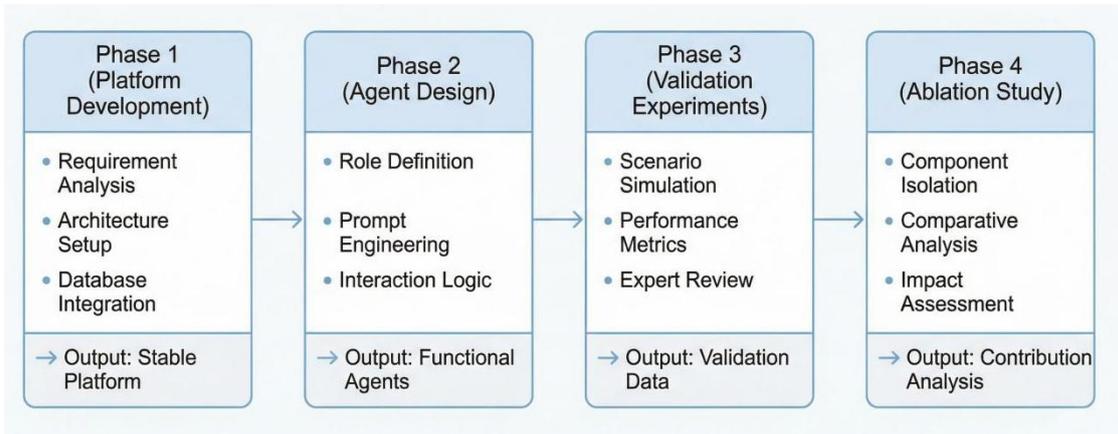

Figure 1: Overview of the four-phase study design for ELDER-SIM development and validation.

## 2.1 Platform Architecture

ELDER-SIM was developed as a modular, microservices-oriented platform to support reproducible construction and evaluation of LLM-based elderly digital twin agents across heterogeneous care contexts. The platform comprises 5 functional layers: (1) workflow orchestration, (2) LLM inference, (3) agent management, (4) memory systems, and (5) evaluation modules (Figure 2). Together, these layers separate scenario control (workflows), response generation (LLMs), identity constraints (agent profiles and cognitive models), persistence (memory), and measurement (psychometric outputs), enabling controlled experiments under standardized conditions.

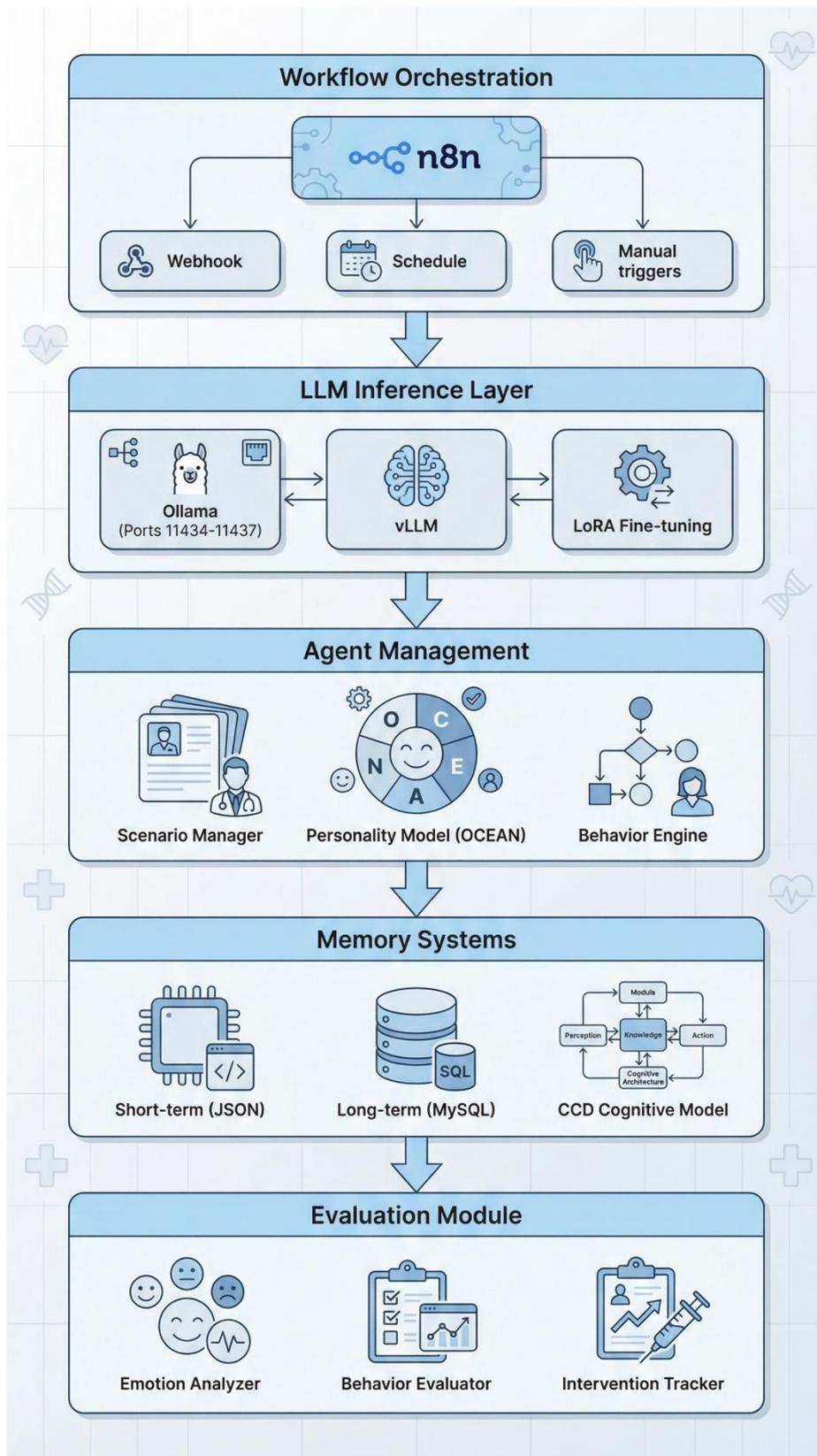

Figure 2: Multi-layer architecture of the ELDER-SIM platform showing the integration of workflow orchestration, LLM inference, agent management, memory systems, and evaluation modules.

### 2.1.1 Workflow Orchestration Layer

Multi-agent interactions were orchestrated using n8n (version 1.0+), which provides a visual workflow engine and supports high-complexity automation through extensible integrations. ELDER-SIM implemented four reusable workflow patterns: (1) a dual-agent dialogue pattern for turn-based older adult–clinician conversations; (2) a multi-agent social simulation pattern supporting event-driven interactions among older adults, family members, clinicians, and community actors; (3) a structured intervention protocol pattern implementing multi-round, evidence-based sequences (eg, 6-round cognitive behavioral therapy workflows); and (4) an assessment protocol pattern that standardizes scenario prompts for psychometric testing (Table 1).

Table 1: Workflow Patterns Implemented in ELDER-SIM

| Pattern | Agents | Interaction Type | Use Case |
| --- | --- | --- | --- |
| Dual-Agent dialogue | 2 (older adult + provider) | Turn-based, dyadic conversation | Routine clinical encounter simulation (eg, symptom review, counseling, medication discussion) |
| Multi-Agent social simulation | 3-6 (older adult + family + community) | Event-driven, multi-party interactions | Social context modeling (eg, loneliness, family conflict, community support, discrimination exposure) |
| Structured intervention protocol | 2 (older adult + therapist) | Protocolized, multi-round sessions | Intervention delivery workflow (eg, 6-session CBT-based sequence with prespecified goals per session) |
| Assessment Protocol | 2 (older adult + evaluator) | Standardized prompts with repeated administrations | Personality consistency evaluation and benchmarking (eg, repeated scenario runs for psychometric scoring and role discrimination) |

### 2.1.2 LLM Inference Layer

ELDER-SIM supported multiple inference backends to accommodate different deployment constraints. For local inference, the platform used an Ollama backend with concurrent instances to enable parallel processing of scenarios; the default model was Qwen2.5-14B. For higher-throughput use cases, vLLM was integrated via an OpenAI-compatible API, enabling efficient batch processing with tensor parallelism. Inference parameters were prespecified and held constant across experimental conditions (eg, temperature=0.7; top-p=0.9; maximum tokens=512; context window=8192; 4 GPU instances; Q8_0 quantization). Table 2 presents the LLM configuration parameters used in our experiments.

Table 2: LLM Configuration Parameters

| Parameter | Value | Description |
| --- | --- | --- |
| Model | Qwen2.5-14B | Base model for agent responses |
| Temperature | 0.7 | Controls response randomness |
| Top-p | 0.9 | Nucleus sampling threshold |
| Max tokens | 512 | Maximum response length |
| Context window | 8192 | Maximum context length |
| GPU instances | 4 | Parallel processing capacity |
| Quantization | Q8_0 | Memory-efficient quantization |

### 2.1.3 Fine-Tuning Module (Domain Adaptation)

To improve domain alignment and trait stability, we implemented low-rank adaptation (LoRA) fine-tuning using the unsloth framework. The base model was Qwen2.5-7B-Instruct (4-bit quantization), with LoRA rank $r=16$ and $\alpha=16$ applied to attention and feed-forward projection modules. Training data comprised 19,717 instruction pairs derived from the China Health and Retirement Longitudinal Study (CHARLS), converted into conversational formats representing older adults' perspectives. Training was conducted for 2000 steps (batch size=2; gradient accumulation=4; learning rate=$2\times10^{-4}$; warmup ratio=0.03; AdamW 8-bit), achieving a final training loss of approximately 0.05. The fine-tuned model was exported to GGUF format and deployed through Ollama with Q8_0 quantization.

## 2.2 Agent Design

### 2.2.1 OCEAN Personality Model

Each agent in ELDER-SIM was characterized using the Big Five (OCEAN) personality model [27, 28]. This framework comprises five orthogonal dimensions—openness, conscientiousness, extraversion, agreeableness, and neuroticism—that have been extensively validated across cultures and populations [29]. We selected OCEAN to support the development of personality-stable elderly digital twins, because longitudinal credibility requires trait-consistent patterns in language, interpersonal stance, and coping responses across repeated interactions rather than plausibility in a single encounter. For implementation, each OCEAN trait was prespecified as a numeric target on a 1-to-5 scale and incorporated into the agent's generation context to constrain tone, behavioral tendencies, and decision-making style. Operational definitions and behavioral indicators for high vs low trait expression are summarized in Table 3.

Table 3: Detailed OCEAN Personality Dimensions with Behavioral Indicators

| Dimension | High Score Indicators | Low Score Indicators | Scale |
|---|---|---|---|
| Openness (O) | Curious, creative, open to new experiences, imaginative | Conventional, practical, prefers routine, traditional | 1–5 |
| Conscientiousness (C) | Organized, dependable, self-disciplined, goal-oriented | Careless, impulsive, disorganized, flexible | 1–5 |
| Extraversion (E) | Sociable, assertive, energetic, talkative | Reserved, quiet, solitary, reflective | 1–5 |
| Agreeableness (A) | Cooperative, trusting, helpful, empathetic | Competitive, skeptical, challenging, detached | 1–5 |
| Neuroticism (N) | Anxious, moody, emotionally reactive, vulnerable | Calm, stable, resilient, secure | 1–5 |

### 2.2.2 Agent Configuration Schema

Agent identity and context in ELDER-SIM were specified using structured JSON configuration files to enable reproducible instantiation of elderly digital twin agents across scenarios and experimental conditions. Each agent profile followed a standardized schema capturing (1) demographics (eg, age, gender, education, occupation, living situation, marital status, children, income level), (2) health status (eg, chronic conditions, cognitive status, functional status, medication adherence), (3) personality parameters (OCEAN trait values on a 1–5 scale), (4) social context (eg, family support, community engagement, discrimination history), and (5) behavioral constraints expressed as natural-language rules that operationalized stable communication habits and role-consistent mannerisms (eg, response pacing, preferred topics, value orientation, reluctance to disclose emotions, or distrust of technology). These profile elements were injected into the agent's generation context to constrain tone, interpersonal stance, and decision tendencies while allowing content to vary appropriately with the scenario prompt.

**1.1.1** To support extensibility across care settings, the schema was designed to separate relatively stable attributes (eg, demographics and personality parameters) from context-sensitive attributes (eg, current health concerns, social stressors, and scenario-triggered behavioral constraints), enabling controlled manipulation of scenario inputs without altering the underlying digital twin specification. An example JSON profile is provided in Listing 1, illustrating how a single agent configuration can encode person-level characteristics alongside clinically relevant contextual modifiers.

Listing 1: Example Agent Configuration for Older Adults

```
{
  "agent_id": "elderly_patient_001",
  "demographics": {
    "name": "Wang Daye",
    "age": 72,
    "gender": "male",
    "education": "high_school",
    "occupation": "retired_teacher",
    "living_situation": "living_alone",
    "marital_status": "widowed",
    "children": 2,
    "income_level": "moderate"
  },
  "health_status": {
    "chronic_conditions": ["hypertension", "type2_diabetes"],
    "cognitive_status": "normal",
    "functional_status": "independent",
    "medication_adherence": "moderate"
  },
  "personality":
    { "openness": 2.5,
    "conscientiousness": 4.0,
    "extraversion": 2.0,
    "agreeableness": 3.5,
    "neuroticism": 4.0
  },
  "social_context":
    { "family_support": "low",
    "community_engagement": "minimal",
    "discrimination_history": ["workplace", "healthcare"]
  },
  "behavior_constraints": [
    "Speaks slowly and deliberately",
    "Often references past experiences as a teacher",
    "Shows reluctance to discuss emotional topics",
    "Demonstrates traditional values",
    "Expresses distrust of modern technology"
  ]
}
```

### 2.2.3 Role Types and Characteristics

ELDER-SIM is a multi-role platform in which each agent is instantiated with a predefined role type to ensure role-consistent behavior across elderly-care scenarios. Role definitions specify the agent's interaction mandate (eg, older adult self-report vs professional assessment), communication stance, and expected behavioral patterns, while preserving the assigned personality configuration. Role types included older adults, health care providers, family members, community actors, and institutional/service roles. This structure enables standardized scenario prompts to be applied consistently while eliciting role-appropriate responses across experimental conditions (Table 4).

Table 4: Role Types and Characteristics in ELDER-SIM

| Category | Roles | Key Characteristics | Typical OCEAN |
|---|---|---|---|
| Older adults | Solitary older adult; chronic disease adult; cognitive impairment adult | Primary simulation targets representing heterogeneous health and social conditions | Low E, high N |
| Health care providers | Geriatrician; nurse; rehabilitation therapist | Professional communication and evidence-based practice orientation | High C, high A |
| Family members | Supportive child; neglectful child; | Variable support levels with emotionally salient | Variable |

| | discriminatory child; spouse | relationship dynamics | |
| --- | --- | --- | --- |
| Community | Social worker; neighbor; stranger | Social interaction patterns, including potential sources of discrimination | Variable |
| Institutional | Nursing home staff; social services | Formal care provision shaped by organizational processes and policy constraints | High C |

*OCEAN indicates the Big Five personality dimensions: O (openness), C (conscientiousness), E (extraversion), A (agreeableness), N (neuroticism).

### 2.3 Memory System

ELDER-SIM implements a 3-tier memory architecture to support longitudinal agent behavior and to reduce within-agent drift during repeated interactions. The design was informed by human memory models and cognitive psychology research [30, 31] and separates short-horizon conversational context from durable, person-level memory traces that can be retrieved across sessions.

#### 2.3.1 Short-Term Memory

Short-term memory maintained the immediate conversational context using a sliding-window mechanism. For fast read/write access during inference, short-term memory was stored as JSON records with timestamps and metadata. The default capacity was up to 100 dialogue turns (configurable). Stored content included (1) dialogue history with speaker attribution, (2) a current emotional-state vector (anxiety, fear, sadness, anger, shame, guilt, and happiness), (3) active automatic thoughts triggered by recent events, and (4) a running conversation summary to compress context while preserving salient information. Updates followed a first-in–first-out rule with importance-based retention to preferentially preserve clinically salient content.

#### 2.3.2 Long-Term Memory (MySQL)

Long-term memory was persisted in a MySQL database (version 8.0+) using a normalized schema to support efficient retrieval and relationship tracking (Table 5). The schema distinguished: (1) episodic memory (time-stamped events with emotional valence and importance), (2) semantic memory (factual knowledge and beliefs with confidence and source), (3) belief updates (tracked changes in core and intermediate beliefs, including the trigger event and timestamp), and (4) dialogue summaries (session-level compressed histories with key topics and emotional trajectory). This structure enabled agents to carry forward clinically relevant life events (eg, diagnoses, hospitalizations, family conflict, discrimination experiences) and stable psychosocial context for subsequent interactions.

Table 5: Long-Term Memory Database Schema

| Table | Primary key | Core fields | Purpose | Example content |
| --- | --- | --- | --- | --- |
| episodic_memory | memory_id | agent_id, event_type, event_time, content, emotional_valence, importance, metadata_json | Stores time-stamped personal episodes that shape the agent's narrative identity and subsequent responses. | "Hospitalized for pneumonia (2021-11); felt fearful; importance=0.9" |
| semantic_memory | fact_id | agent_id, category, content, confidence, source, updated_at | Stores stable factual knowledge and preferences (health beliefs, routines, resources), enabling consistent recall. | "Believes antihypertensives cause dizziness; confidence=0.7; source=caregiver" |
| belief_updates | update_id | agent_id, belief_level, belief_type, old_value, | Logs changes in core/intermediate | "Core belief: 'I'm a burden' → 'I can |

| | | new_value, trigger_event, timestamp | beliefs over time for cognitive consistency tracking and auditability. | still contribute' after supportive visit" |
| --- | --- | --- | --- | --- |
| dialogue_summaries | summary_id | session_id, agent_id, summary, key_topics_json, emotional_trajectory_json, created_at | Stores compressed session memories for long-horizon continuity without bloating the prompt context. | Summary + key topics (medication, loneliness) + emotion curve (sad →calm) |

### 2.3.3 Cognitive Conceptualization Diagram (CCD)

Because memory persistence alone does not specify how past experiences shape appraisal and behavior, ELDER-SIM incorporated a Cognitive Conceptualization Diagram (CCD) module based on Beck's cognitive model [25]. The CCD operationalized a structured pathway linking (1) background vulnerabilities and relevant history, (2) belief systems (core beliefs, intermediate beliefs, and coping strategies), and (3) situation-evoked outputs (automatic thoughts, emotions, and behaviors). In this framework, a situation trigger activates automatic thoughts constrained by the belief architecture; emotions are represented as intensity values on a 0–1 scale, and behaviors are generated as observable manifestations of the cognitive–emotional process (Table 6). By anchoring responses to an explicit, theory-consistent structure, the CCD was intended to improve cognitive coherence and stabilize trait-consistent behavior across repeated encounters.

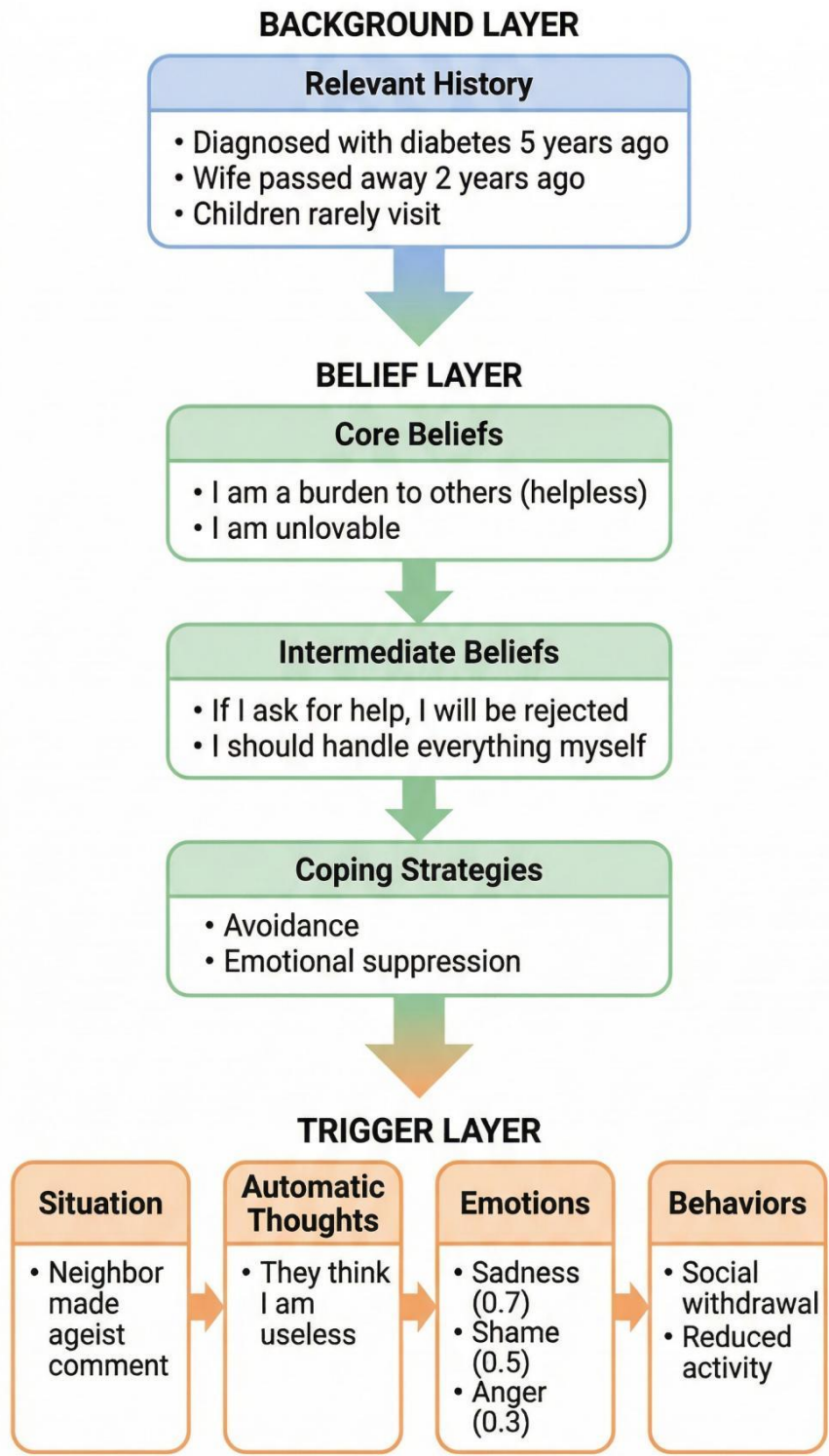

Figure 3: Three-layer Cognitive Conceptualization Diagram (CCD) architecture showing the relationship between background history, belief systems, and triggered responses.

Table 6: CCD Components and Descriptions

| Layer | Component | Operational definition (for agent modeling) | Typical representation |
|---|---|---|---|
| Background | Relevant history | Salient past experiences that calibrate vulnerability/meaning- | Key life events, losses, diagnoses, trauma, discrimination |
| | | making and inform schema activation. | experiences |
| Belief system | Core beliefs | Deep, rigid schemas about self/others/world that bias interpretation across contexts. | "I am helpless / unlovable / worthless" (schema labels + strength) |
| | Intermediate beliefs | Conditional rules/assumptions derived from core beliefs; guide choices under uncertainty. | "If I ask for help, people will reject me" |
| | Coping strategies | Habitual behavioral/cognitive strategies used to manage distress (often maintaining the schema). | Avoidance, reassurance seeking, suppression, over-compliance |
| Trigger/Context | Triggering situation | External event/context that activates the belief system and elicits appraisals. | "Child criticized medication adherence" / "Nurse used impatient tone" |
| | Automatic thoughts | Rapid, situation-linked interpretations that mediate emotion and behavior. | Short evaluative statements + intensity/credibility |
| Response | Emotions | Affective responses produced by appraisals; can be modeled as multi-dimensional intensities. | Emotion vector (e.g., anxiety/sadness/anger 0–1) |
| | Behaviors | Observable actions/communication styles generated as outputs of the cognitive-emotional process. | Refusal, withdrawal, irritability, help-seeking, compliance, disclosure |

## 2.4 Psychometric Evaluation Framework and Experimental Design

To evaluate whether LLM−based agents can serve as clinically credible, person-consistent building blocks for elderly-care digital twins, we adapted established psychometric methods from personality psychology to quantify personality consistency and role separability across repeated administrations and system configurations [20, 28, 32].

### 2.4.1 Reliability Measures

**Internal Consistency (Cronbach's $\alpha$):** Internal consistency was quantified using Cronbach α, which estimates the extent to which items intended to measure the same latent construct produce coherent scores within a scale. Higher values indicate greater within-scale coherence. We interpreted α using conventional thresholds (excellent, ≥ 0.90; good, 0.80–0.89; acceptable, 0.70–0.79; questionable, 0.60–0.69; poor, <0.60).

**Test-Retest Reliability (ICC):** Stability across repeated administrations was assessed using ICC for absolute agreement. We interpreted ICC using standard cut points (excellent, ≥0.90; good, 0.75–0.89; moderate, 0.50–0.74; poor, <0.50).

### 2.4.2 Validity Measures

**Role Discrimination Accuracy:** Because digital-twin use cases require that distinct agent roles and personality configurations remain detectably different (ie, not collapsing into generic responses), we evaluated role discrimination using classification accuracy derived from true-positive, true-negative, false-positive, and false-negative counts. Higher accuracy indicates better separation between agent profiles under identical scenario prompts.

### 2.4.3 Experimental Design

**Scenario Generation:** We developed 10 standardized elderly-care scenarios designed to elicit personality-relevant behavioral signals across common, clinically salient contexts (eg, medication adherence, family conflict, social isolation, health anxiety, financial concerns, grief, ageism, care transitions, technology frustration, and end-of-life planning)

Table 7: Standardized Test Scenarios for Personality Assessment

| ID | Scenario | Description |
|---|---|---|
| S1 | Medication Adherence | The older adult discusses reluctance to take prescribed medications. |
| S2 | Family Conflict | The older adult describes disagreement with adult children about. |
| S3 | Social Isolation | The older adult expresses feelings of loneliness and disconnection. |
| S4 | Health Anxiety | The older adult worries about worsening health condition. |
| S5 | Financial Concerns | The older adult discusses worries about healthcare costs. |
| S6 | Loss and Grief | The older adult processes recent loss of spouse or friend. |
| S7 | Ageism Experience | The older adult recounts experience of age discrimination. |
| S8 | Care Transition | The older adult discusses moving to assisted living facility. |
| S9 | Technology Frustration | The older adult expresses difficulty with digital health tools. |
| S10 | End-of-Life Planning | The older adult discusses advance care preferences. |

Note: Each scenario was administered 5 times per agent configuration; responses were scored on OCEAN dimensions using a 1–5 scale and used to compute reliability and validity metrics.

**Assessment Protocol:** For each agent configuration, we presented standardized prompts containing demographic and personality context and administered each of the 10 scenarios 5 times (50 total scenario administrations per agent). Responses were evaluated using a rubric to extract behavioral indicators and to assign OCEAN (Big Five) dimension scores on a 1-5 scale, after which reliability and validity metrics were computed.

**Ablation Conditions:** To isolate the contribution of specific system components to personality consistency—an essential property for longitudinal digital-twin fidelity—we compared 4 incrementally enhanced configurations: (1) Baseline (prompt-only personality description), (2) +Memory (short-term and long-term memory), (3) +CCD (+Memory plus Cognitive Conceptualization Diagram), and (4) +LoRA (+CCD plus domain-specific fine-tuning).

## 3. Results

### 3.1 Participant and Data Characteristics

We generated personality assessment data across four experimental conditions using 6 distinct agent configurations (3 older adult profiles and 3 healthcare provider profiles). Each agent was administered 10 standardized scenarios with 5 repetitions, yielding 300 responses per condition (1,200 total responses across all conditions). Across conditions, the number of agent configurations, scenarios, repetitions, and total responses were held constant (6 configurations, 10 scenarios per configuration, 5 repetitions per scenario, and 300 responses per condition). Mean response length increased from 156.3 tokens in the Baseline condition to 178.2 tokens in the +Memory condition and 203.7 tokens in the +CCD condition, and was 189.4 tokens in the +LoRA condition. Mean response time ranged from 1234 ms (Baseline) to 1678 ms (+CCD). Table 8 presents the data characteristics for each experimental condition.

Table 8: Data Characteristics by Experimental Condition

| Characteristic | Baseline | +Memory | +CCD | +LoRA |
|---|---|---|---|---|
| Agent configurations | 6 | 6 | 6 | 6 |
| Scenarios per agent | 10 | 10 | 10 | 10 |
| Repetitions per scenario | 5 | 5 | 5 | 5 |
| Total responses | 300 | 300 | 300 | 300 |
| Mean response length (tokens) | 156.3 | 178.2 | 203.7 | 189.4 |
| Response time (ms) | 1,234 | 1,456 | 1,678 | 1,523 |

## 3.2 Personality Consistency Validation

### 3.2.1 Internal Consistency

Table 9 presents Cronbach's $\alpha$ values for each OCEAN dimension across experimental conditions. All conditions achieved acceptable internal consistency ($\alpha > 0.70$). The +CCD condition showed substantial improvement over Baseline and +Memory conditions, with mean $\alpha$ increasing from 0.70 to 0.89 (27.1% improvement). The +LoRA condition achieved excellent internal consistency ($\alpha = 0.94$).

Table 9: Internal Consistency (Cronbach's $\alpha$) by Condition and Personality Dimension

| Condition | O | C | E | A | N | Mean (SD) |
|---|---|---|---|---|---|---|
| Baseline | 0.68 | 0.72 | 0.71 | 0.70 | 0.70 | 0.702 (0.015) |
| +Memory | 0.69 | 0.73 | 0.70 | 0.71 | 0.70 | 0.705 (0.013) |
| +CCD | 0.88 | 0.91 | 0.89 | 0.90 | 0.88 | 0.892 (0.013) |
| +LoRA | 0.93 | 0.95 | 0.94 | 0.95 | 0.93 | 0.940 (0.010) |

### 3.2.2 Test-Retest Reliability

All conditions demonstrated good to excellent test-retest reliability (ICC > 0.85), indicating that personality measurements were stable across repeated administrations. The +CCD and +LoRA conditions achieved excellent reliability (ICC > 0.90). Table 10 presents ICC values for repeated personality assessments across conditions.

Table 10: Test-Retest Reliability (ICC) by Condition

| Condition | ICC | 95% CI | F-statistic | Interpretation |
|---|---|---|---|---|
| Baseline | 0.856 | [0.79, 0.91] | 12.9 | Good |
| +Memory | 0.871 | [0.81, 0.92] | 14.5 | Good |
| +CCD | 0.924 | [0.87, 0.96] | 25.3 | Excellent |
| +LoRA | 0.958 | [0.92, 0.98] | 46.2 | Excellent |

### 3.2.3 Role Discrimination

The Baseline condition showed moderate role discrimination (83.3%), with occasional confusion between agents with similar personality profiles. Enhanced conditions showed progressive improvement: +Memory (88.9%), +CCD (94.4%), and +LoRA (97.2%), demonstrating the cumulative benefits of memory, cognitive modeling, and domain-specific fine-tuning. Table 11 presents role discrimination metrics for distinguishing between different agent personality profiles.

Table 11: Role Discrimination Metrics by Condition

| Condition | Accuracy | Precision | Recall | F1 | AUC |
|---|---|---|---|---|---|
| Baseline | 83.3% | 0.82 | 0.83 | 0.82 | 0.89 |
| +Memory | 88.9% | 0.88 | 0.89 | 0.88 | 0.93 |
| +CCD | 94.4% | 0.94 | 0.94 | 0.94 | 0.97 |
| +LoRA | 97.2% | 0.97 | 0.97 | 0.97 | 0.99 |

## 3.3 Ablation Study

To quantify the incremental contribution of key system components to personality consistency, we compared 4 prespecified configurations under otherwise identical testing conditions: Baseline (prompt-only personality description), +Memory (short-term and long-term memory), +CCD (+Memory plus Cognitive Conceptualization Diagram), and +LoRA (+CCD plus domain-specific fine-tuning).

### 3.3.1 Component Contributions

Table 12 summarizes the ablation study results. Across configurations, personality consistency improved in a stepwise manner. Mean internal consistency (Cronbach's $\alpha$) was 0.702 in Baseline and 0.706 with +Memory, increased to 0.892 with +CCD, and reached 0.940 with +LoRA. Similar monotonic improvements were observed for test-retest reliability (ICC: 0.856, 0.871, 0.924, and 0.958, respectively), role discrimination accuracy (83.3%, 88.9%, 94.4%, and 97.2%), and response coherence (3.2/5, 3.5/5, 4.6/5, and 4.8/5)

Table 12: Ablation Study Summary: Component Contributions to Personality Consistency

| Metric | Baseline | +Memory | +CCD | +LoRA |
|---|---|---|---|---|
| Cronbach's $\alpha$ | 0.702 | 0.706 | 0.892 | 0.940 |
| $\Delta\alpha$ from Baseline | – | +0.004 | +0.190 | +0.238 |
| ICC | 0.856 | 0.871 | 0.924 | 0.958 |
| Role Discrimination | 83.3% | 88.9% | 94.4% | 97.2% |
| Response Coherence | 3.2/5 | 3.5/5 | 4.6/5 | 4.8/5 |

### 3.3.2 Statistical Analysis

For the primary ablation contrast on internal consistency (Table 13), we conducted paired $t$ tests with Bonferroni correction. The increase in $\alpha$ from Baseline to +Memory was small and not statistically significant ($\Delta\alpha$ =0.004; $p$=1.000). In contrast, +CCD was associated with a statistically significant increase relative to Baseline ($\Delta\alpha$ =0.190; $p$<0.001) and relative to +Memory ($\Delta\alpha$=0.186; $p$<0.001). +LoRA yielded an additional increase over +CCD ($\Delta\alpha$=0.048; $p$=0.028).

Table 13: Pairwise Comparisons of Cronbach's $\alpha$ (Bonferroni-corrected)

| Comparison | $\Delta\alpha$ | t | df | p |
|---|---|---|---|---|
| Baseline vs +Memory | +0.004 | 0.52 | 4 | 1.000 |
| Baseline vs +CCD | +0.190 | 6.83 | 4 | <0.001*** |
| Baseline vs +LoRA | +0.238 | 9.21 | 4 | <0.001*** |
| +Memory vs +CCD | +0.186 | 6.56 | 4 | <0.001*** |
| +Memory vs +LoRA | +0.234 | 8.98 | 4 | <0.001*** |
| +CCD vs +LoRA | +0.048 | 2.87 | 4 | 0.028* |

Note: *$p < 0.05$, **$p < 0.01$, ***$p < 0.001$

### 3.3.3 Dimension-Specific Analysis

Table 14 presents the improvement in Cronbach's $\alpha$ for each OCEAN dimension. In the Baseline condition, Cronbach's $\alpha$ ranged from 0.68 (Openness) to 0.72 (Conscientiousness). With +CCD, $\alpha$ increased to 0.88 − 0.91 across dimensions (Openness, 0.88; Conscientiousness, 0.91; Extraversion, 0.89; Agreeableness, 0.90; Neuroticism, 0.88), corresponding to absolute gains of 0.18 to 0.20. With +LoRA, $\alpha$ increased further to 0.93 − 0.95 (Openness, 0.93; Conscientiousness, 0.95; Extraversion, 0.94; Agreeableness, 0.95; Neuroticism, 0.93), with total gains vs Baseline of 0.23 to 0.25.

Table 14: Dimension-Specific Improvements in Internal Consistency

| Dimension | Baseline | +CCD | +LoRA | $\Delta$ (CCD) | $\Delta$ (LoRA) |
|---|---|---|---|---|---|
| Openness | 0.68 | 0.88 | 0.93 | +0.20 | +0.25 |
| Conscientiousness | 0.72 | 0.91 | 0.95 | +0.19 | +0.23 |
| Extraversion | 0.71 | 0.89 | 0.94 | +0.18 | +0.23 |
| Agreeableness | 0.70 | 0.90 | 0.95 | +0.20 | +0.25 |
| Neuroticism | 0.70 | 0.88 | 0.93 | +0.18 | +0.23 |

## 4. Discussion

In this study, we developed ELDER-SIM, a multi-role elderly-care simulation platform and proposed a psychometric framework to quantify personality stability of LLM-based elderly digital twin agents. Under a standardized, repeated-scenario testing design, we found that personality consistency was achievable at clinically meaningful reliability levels when identity constraints were implemented with theory-driven cognitive structure and domain adaptation. Specifically, internal consistency improved from acceptable levels in the prompt-only baseline to excellent levels with CCD and LoRA, accompanied by excellent test–retest reliability and high role discrimination. These results indicate that "personality drift" is not an unavoidable property of LLM agents; rather, it is a modifiable system behavior that can be reduced using constrained architectures and measurable reliability targets.

A central barrier to clinically credible elderly digital twins is personality drift across repeated encounters. Our findings provide empirical evidence that personality consistency is achievable in LLM-based agents under a controlled, repeated-administration design. Across all configurations—including the prompt-only baseline—internal consistency met conventional acceptability thresholds, and test-retest reliability was good to excellent. These results are important because they indicate that, even without extensive system augmentation, trait expression in LLM agents is not purely noise; rather, it can be captured as a measurable construct with psychometric properties. This aligns with emerging psychometric evaluations of LLM personality, which show that instruction-tuned models can yield coherent trait measurements when assessed systematically [19,20,33]. In geriatric psychosocial care contexts—where longitudinal continuity of coping style, interpersonal stance, and health-related behaviors is clinically consequential [4-7]—establishing that trait stability is measurable and at least acceptable even in baseline settings is a prerequisite for any credible "trajectory" or "intervention-response" simulation. The practical significance is reproducibility: if personality scores are not stable, simulated trajectories cannot support meaningful scenario testing. Demonstrating acceptable baseline reliability provides a minimum viable foundation for safe use in educational rehearsal and early-stage intervention prototyping, while also defining quantitative targets for iterative system improvement [32,34,35].

The most consequential result in this study is that cognitive modeling produced the largest gain in personality consistency. Adding CCD increased mean Cronbach α from approximately 0.70 to 0.89, representing a substantial improvement in within-scale coherence. This finding supports the hypothesis that structured cognitive frameworks are essential for realistic and stable personality simulation. Conceptually, personality-consistent behavior over repeated encounters requires more than recalling prior dialogue; it requires a stable mechanism linking "what happened" to "how the agent appraises it" and "how it responds." The CCD module operationalizes this pathway using Beck's cognitive model—background vulnerabilities and history, belief systems (core/intermediate beliefs and coping strategies), and situation-triggered automatic thoughts/emotions/behaviors [25]. By constraining response generation through an explicit belief-appraisal-behavior structure, CCD likely reduces degrees of freedom that otherwise allow the model to "reinterpret" the same scenario inconsistently across runs, thereby reducing drift. This mechanism-level interpretation is consistent with work emphasizing cognitive scaffolds for simulated patients (eg, PATIENT-Ψ), where cognitive conceptualization improves perceived fidelity [10]. Our study extends this literature by demonstrating that cognitive structuring improves not only qualitative realism but also quantitative psychometric reliability, which is more directly relevant for longitudinal digital twin use cases. For clinical-grade elderly digital twins, the relevant design principle is that identity stability is a cognitive-architecture problem, not a prompt-formatting problem. CCD-like scaffolds may therefore be necessary when digital twins are intended for repeated interactions (eg, multi-session counseling, adherence support, family conflict mediation), in which drift would otherwise undermine trust and interpretability.

Beyond stability, clinical simulation requires that trait expression be behaviorally plausible and context-aligned, especially for older adults. In our ablation, LoRA fine-tuning achieved the highest reliability indices while also improving the qualitative naturalness of personality expression. A plausible interpretation is that domain adaptation improves the model's alignment with elderly-care discourse patterns and contextual priors (eg, how older adults describe symptoms, social stressors, technology frustration, or care transitions), thereby making trait-consistent responses easier to sustain. This complements prior evidence that instruction tuning improves personality measurability [20,33] and that domain-specific modeling can enhance clinical appropriateness of LLM behavior [36]. Importantly, the incremental gain of LoRA over CCD suggests that fine-tuning is most effective once a stabilizing cognitive scaffold is in place—ie, LoRA may refine and "densify" trait-consistent language and behaviors rather than replace the need for structured cognition. For deployment-oriented simulation, fine-tuning can be viewed as a fidelity layer: after cognitive stability is achieved, LoRA can improve the perceived authenticity of older-adult narratives and clinician–patient dynamics, potentially enhancing educational transfer and face validity for intervention prototyping. However, because fine-tuning can also change behavioral distributions, psychometric auditing should remain part of the post–fine-tuning validation pipeline to ensure that gains in naturalness do not come at the expense of unwanted bias or unsafe responses [23,37,38].

A striking and practically relevant negative finding is that memory persistence alone did not materially improve personality consistency. This result challenges an increasingly common assumption in LLM agent engineering: that adding long-term memory is the primary route to stable identity. From a cognitive standpoint, this is unsurprising—episodic recall supports narrative continuity, but does not specify the stable rules that govern appraisal and coping [30,31]. Without an explicit belief–appraisal mechanism, memory can even introduce additional variability by providing more retrieval candidates and alternative interpretations, increasing the chance that the agent will "shift stance" across repeated administrations. In other words, memory can preserve content while still allowing drift in style, preference, or coping tendencies. In geriatric simulation, implementing memory without cognitive constraints risks producing agents that remember facts but behave inconsistently—an undesirable combination for clinical training and any prospective in silico testing. Practically, this argues for an architecture in which memory is subordinated to a stable cognitive/personality kernel (eg, CCD + trait constraints), rather than used as the primary stabilizer.

### 4.1 Limitations

This study has several limitations. First, the findings were generated using Qwen-based models and a specific local deployment stack; personality stability and role separability may differ across other model families, decoding settings, and serving infrastructures. Second, the platform and fine-tuning data were anchored in Chinese elderly-care contexts, with adaptation data derived from CHARLS; generalizability to other cultural and health-system settings will require revalidation using locally appropriate corpora and scenario designs. Third, evaluation was conducted entirely in simulated interactions. Although we demonstrated strong psychometric reliability, we did not test whether the simulated trajectories align with real older adults' longitudinal behaviors, clinician assessments, or clinical outcomes, and ecological validity therefore remains uncertain. Fourth, personality scoring relied on rubric-based ratings of generated responses; we did not incorporate blinded multi-rater human coding or external criterion measures of personality, which may introduce measurement error or systematic bias in trait estimation. Fifth, the standardized battery included 10 scenarios with repeated administrations; broader scenario coverage, longer time horizons, and more diverse interactional contexts may yield different stability profiles, particularly for edge cases such as cognitive impairment, multimorbidity, or crisis situations. Finally, the most stable configuration required substantial computational resources, which may limit reproducibility for some groups and highlights the need for more efficient stability-preserving designs.

## 5. Conclusion

In this study, ELDER-SIM demonstrated that personality-stable elderly digital twin agents are technically achievable and measurable under repeated scenario administration. Across controlled ablations, cognitive structuring with a Cognitive Conceptualization Diagram produced the largest improvement in personality consistency, domain adaptation with LoRA further increased reliability and role separability, and memory persistence alone yielded minimal incremental benefit. Together, these findings suggest that clinically credible longitudinal simulation will likely require theory-informed cognitive constraints coupled with domain-aligned adaptation, evaluated against prespecified psychometric reliability targets. Future research should validate these agents against real-world longitudinal encounters across diverse cultural and care settings, expand scenario coverage and time horizons, and examine whether stability gains translate into improved educational effectiveness, safer deployment, and more reliable in silico testing of geriatric psychosocial interventions.

## Acknowledgments

None.

## Conflicts of Interest

The authors declare no conflicts of interest.

## Data Availability

The ELDER-SIM platform code, agent configurations, and evaluation scripts are available at [repository URL]. The CHARLS-derived fine-tuning dataset is available upon reasonable request.